\documentclass[aps,amsfonts,nofootinbib,twocolumn]{revtex4}

\usepackage{hyperref}
\usepackage{wasysym}
\usepackage{graphicx}

\def\[{\left\lbrack}
\def\]{\right\rbrack}
\def\({\left(}
\def\){\right)}
\newcommand{\be}{\begin{equation}}
\newcommand{\ee}{\end{equation}}
\newcommand{\ea}{\end{eqnarray}}
\newcommand{\ba}{\begin{eqnarray}}

\newcommand{\asp}{\textquotedblleft}

\newcommand{\diag}{\mbox{diag}}

\newcommand{\flrw} {{\mbox{\tiny RW$\tilde \Lambda$}}}
\newcommand{\dec} {{\mbox{\tiny dec}}}

\begin{document}

\title{Anisotropic Cosmological Constant and the CMB Quadrupole Anomaly}

\author{Davi C. Rodrigues}
\email{drodrigues@fisica.usach.cl}
\affiliation{Departamento de F{\'\i}sica,
Universidad de Santiago de Chile, Casilla 307, Santiago, Chile }

\begin{abstract}
    There are evidences that the cosmic microwave background (CMB) large-angle anomalies imply a departure from statistical isotropy and hence from the standard cosmological model. We propose a $\Lambda$CDM model extension whose dark energy component preserves its nondynamical character but wield anisotropic vacuum pressure. Exact solutions for the cosmological scale factors are  presented, upper bounds for the deformation parameter  are evaluated and its value is estimated considering the elliptical universe proposal to solve the quadrupole anomaly. This model can be constructed from a Bianchi I cosmology with cosmological constant from two different ways: i) a straightforward anisotropic modification of the vacuum pressure consistently with energy-momentum conservation; ii) a Poisson structure deformation between canonical momenta such that the dynamics remain invariant under scale factors rescalings.\\[0.1in] 
\small{ Keywords: dark energy, CMB, cosmological constant, phantom energy, isotropization.} \\[0.1in] 
PACS numbers: 95.36.+x, 98.70.Vc.
\end{abstract}

\maketitle

\section{Introduction}

Since the observation of the current  cosmic expansion speed up through the light-curve of type Ia Supernovae \cite{super},  which was  confirmed independently by cosmic microwave background (CMB) \cite{cmb_expan, COBE, wmap3} and  large scale structures observations \cite{struct}, it became clear that a pure Friedmann-Lemaitre-Robertson-Walker (FLRW) cosmology with matter and radiation could not explain all the large scale properties of our universe (for reviews see \cite{revde, cmb_rev}). The missing piece was called dark energy and, surprisingly, a single constant, the once despised cosmological constant $\Lambda$,  was capable of describing its behavior with great success; being now part of the standard   $\Lambda$CDM  model \cite{wmap3}. However, the curious smallness of its energy density ($\rho_\Lambda \sim 10^{-47} \mbox{GeV}^4$) and its lack of correspondence with fundamental physics  has raised many questions, and many alternative models have appeared\cite{revde}. While the cosmological constant always leads to an equation of state (EoS) parameter for the vacuum with constant value $-1$ ($\omega =  {p_\Lambda} /{\rho_\Lambda} =-1$), in alternative models its value can vary with time and can be  strictly $\ge -1$ (e.g., quintessence \cite{revde}), strictly $\le -1$ (phantom energy \cite{phantom}) or cross from one region to another depending on time (e.g., quintom \cite{quintom}).  Moreover, the possibility of nonlinear EoS's is also under current research \cite{nonlinear}. Although they are quite different, all these approaches assume that dark energy is isotropic. The possibility of anisotropic dark energy has just started to be evaluated \cite{ans_de, moving, ans_de_eos}.

In this work, motivated by the anomalies found in the CMB anisotropies \cite{wmap3, COBE}, which appear to indicate violation of statistical isotropy \cite{48pol, multve, NS, axisofevil, elliptical, axial_evidence}, and on the increasing interest on Bianchi cosmologies (for a recent review see \cite{revbianchi}), we analyze a nondynamical anisotropic form of dark energy which is generated by the usual cosmological constant $\Lambda$ in combination with a certain parameter  $B$. To this end we employ the Bianchi type I line element as the space-time background. A null $B$ corresponds to the usual cosmological constant and hence leads to the usual isotropic cosmology due to the cosmic no-hair theorem \cite{b_isotropization-w}. Being $B$ non-null, the angular dependence of the Hubble parameter does not fade away as the universe expands; opening the possibility of generating an eccentricity in the last scattering surface \cite{elliptical}. This is achieved without the need of others structures and evades problems with isotropization during inflation or during the current epoch dominated by dark energy. The ``price" of this interesting behavior is violation of the null energy condition (NEC)\footnote{On NEC violation, see \cite{phantom, nec}.}; in particular, this type of dark energy crosses the phantom divide depending on the direction.

Some different approaches to cosmology with different types of noncommutativity  have appeared \cite{nccosmocmb, ncuvir, scalefactornc}, these are commonly motivated by general quantum-gravity expectations and on the possibility of finding traces of it in the CMB anisotropies \cite{nccosmocmb, ncuvir}, which are infrared (IR) consequences of the original ultraviolet (UV) modification \cite{ncuvir}. On the other hand, except for a small fraction of a second, all the universe history is classical and the formation of the last scattering surface occurred when the universe had $2.8 \times 10^{-5}$ of its current age; thus the possibility and consequences of an ``infrared noncommutativity" \cite{infrarednc}, i.e., a direct modification of the classical space-time canonical structure, should also be verified. Here we show that, in the context of a Bianchi I cosmology, the most simple Poisson structure deformation  that does not violate rescaling of the scale factors evolves in time and admits a natural interpretation as describing a new form of dark energy, precisely the one generated by the anisotropic cosmological constant $\Lambda_B$, where $B$ regulates the deformation magnitude.

In the following section we introduce the Poisson structure deformation in the  Bianchi I cosmology, unveiling some of its general features. In Sec. III an axial symmetry is introduced and exact solutions are presented. From the latter, considering the vacuum EoS experimental value and limits on the CMB temperature anisotropies, upper bounds for $|B|$ are found. Consistently with these bounds, the $B$ value is estimated in order to generate an appropriate eccentricity in the last scattering surface capable of solving the quadrupole anomaly \cite{elliptical}.   In the last section we present our conclusions and perspectives.

\section{From Poisson structure deformation to the anisotropic cosmological constant description}
The general relativity action in 4D space-time is given by
\begin{equation}
	S =  \int \[ \frac k 2 ~(R - 2 \Lambda) + {\cal L}_M \] \sqrt{- g} ~ d^4 x,
\end{equation}
where $k = 1 / (8 \pi G)$, $R$ is the Ricci scalar, $\Lambda$ is the cosmological constant, $g$ is the metric determinant and ${\cal L}_M$ the Lagrangian for the matter part.

The FLRW cosmology is an application of general relativity which starts from the very beginning with the assumption of spacial homogeneity and isotropy. In a flat universe this leads to the well known Robertson-Walker metric $ (g_{\mu \nu}) =\diag \pmatrix{ -1 & a^2(t) & a^2(t) & a^2(t)}$.

Bianchi classified a broader class of cosmologies which are homogeneous but not necessarily isotropic \cite{revbianchi}. In the presence of a positive cosmological constant all the nine Bianchi cosmologies (or eight of the nine depending on curvature conditions on the Bianchi IX)  isotropize themselves and become de Sitter spaces asymptotically \cite{b_isotropization-w}. To provide the framework for analyzing cosmological deviations from isotropy, we will use the Bianchi I line element, which reads
\begin{equation}
	ds^2 = - dt^2 + a^2(t) \;  dx^2 + b^2(t) \; dy^2 + c^2(t) \;dz^2.
\end{equation}

We modify the cosmological classical evolution, avoiding late-time isotropization, by deforming the Poisson structure between the scale factors conjugate momenta. This procedure can be pictured as the IR analogue of the  noncommutativity between the scale factors, the latter was explored in  \cite{scalefactornc}. The deformation of the Poisson structure to achieve an  ``IR noncommutativity" has already been explored in other contexts  \cite{infrarednc}. We consider the following relation

\begin{equation}
	\label{nc}
	\{\Pi^i, \Pi^j\}_B =  B^{ij} =  B \; \epsilon^{i j k} \;  a_k,
\end{equation}
where $B$ is a real constant, $i,j,k =1,2,3$, $\epsilon^{1 2 3} = 1$, $a_1 = a$, $a_2 = b$, $a_3 = c$ and $\Pi^i$ is the $a_i$ conjugate  canonical momentum: $\Pi^1 =  -  \tilde k ~ \frac d {dt} (  b  c )$, $\Pi^2 =  -  \tilde k ~ \frac d {dt} (  a  c )$, $\Pi^3 =  -  \tilde k ~ \frac d {dt} ( a  b)$,  where $\tilde k = k ~ \int d^3x$.

Considering as usual $[x^\mu] = m^{-1}$ and $[a_i] = 1$, then $[\Pi^i] = 1$ and $[B] = 1$. On the other hand, for dimensional analyzes it is more useful to adopt the convention $[t] = m^{-1}$, $[x^i] = 1$ and $[a_i] = m^{-1}$, which implies  $[\Pi^i] = m$ and $[B] = m^3$; showing that $B$ should indeed be regarded as an IR deformation. Independently on the convention on dimensions one selects (but preserving $[ds^2] = m^{-2}$), the only possible way to express $B$ in terms of a dimensionless constant $B_0$, without inserting new dimensional parameters, is $B = B_0 ~ \tilde k  ~ \sqrt \Lambda$. In the equations of motion and derivative relations,  $B$ only occurs in the form $B / {\tilde k}$. Considering that $[a_i] = m^{-1}$, $k$ and $\tilde k$ have the same dimension. Hereafter the tilde will not be used.

Eq. (\ref{nc})  should not be interpreted as a Poisson structure deformation induced by   $[\Pi^i, \Pi^j] =  i B^{ij}$, since the latter would insert an improper quantum behavior  into classical observables. Thus, by implementing (\ref{nc}) into the cosmological dynamics, we are not ignoring higher powers of $B$, that is the full deformation.

One important property of  (\ref{nc}) is its invariance by the rescaling of the scale factors [i.e., $a_i \rightarrow a^0_i  ~ a_i$ when $x^i \rightarrow \frac 1 {a^0_i} x^i$ (no sum, the $a^0_i$'s are non-null constants)]. A constant $B^{ij}$, for instance, violates this symmetry. This violation would turn the scale factors into physical observables, in particular it would not be possible to select coordinates such that $a=b=c=1$ at a given time.

The equations of motion of a Bianchi I cosmology with matter, cosmological constant and the Poisson deformation $\{~ ,~ \} \rightarrow \{ ~,~ \}_B = \{~ ,~ \} + B^{ij} ~ \frac \partial {\partial \Pi^i} ~ \frac \partial {\partial \Pi^j}$ read

\begin{eqnarray}
	\frac {\ddot b} b +   \frac {\ddot c} c + H_b \; H_c &=& \Lambda  +  B_0  ~ \sqrt \Lambda ~ \( H_c - H_b  \right) - \frac 1 k T_1^1,\\[.1in]
	\frac {\ddot a} a + \frac {\ddot c}c +H_a \; H_c &=& \Lambda   + B_0 ~ \sqrt \Lambda ~ \left(   H_a  - H_c \right)  - \frac 1 k T_2^2, \\[.1in]
	\frac { \ddot a} a + \frac {  \ddot b} b + H_a \; H_b &=& \Lambda  + B_0 ~ \sqrt \Lambda  \left(  H_b - H_a  \right) - \frac 1 k T_3^3,  \\[.1in]
	 H_a \; H_b + H_a \; H_c &+& H_b \; H_c = \Lambda  - \frac 1 k T_0^0,
\end{eqnarray}
where $T^\nu_\mu (a_i)$ is the matter energy-momentum tensor and $H_{a_i} =\frac { \dot a_i} {a_i}$ is the Hubble parameter in the $x^i$ direction.  The vacuum energy momentum-tensor $({\cal T}_\nu^\mu ) = \diag \pmatrix{ - \rho & p_x & p_y & p_z}$ now depends on $\Lambda$ and $B_0$. The vacuum energy is given by the former, while the anisotropic vacuum pressure depends on both, $B_0$ parametrize the pressure anisotropies.  The EoS parameter is anisotropic, but its mean value is exactly $-1$. \\[.1in]

For the Bianchi I line element, the zero-component of any diagonal energy-momentum tensor divergence is
\begin{equation}
    \label{div0}
	-\nabla_\mu { T}^\mu_0 = \dot \rho + (H_a + H_b + H_c) \rho + H_a ~p_x + H_b ~ p_y + H_c ~p_z.
\end{equation}
This expression is trivially null if $\dot \rho = 0$ and $p_x = p_y=p_z = - \rho$, but it also admits an anisotropic pressure solution,
\begin{eqnarray}
	p_x & = & - \rho +  h~ (H_c - H_b), \nonumber \\
    \label{amomenta}
	p_y & = & - \rho +  h ~(H_a - H_c), \\
	p_z & = & - \rho +  h ~(H_b - H_a), \nonumber
\end{eqnarray}
where $h$ is a constant. Setting $h = - B$, the above pressures are exactly the  vacuum pressures that come from the Bianchi I cosmology with the Poisson deformation. This solution also satisfies $\nabla_\mu {\cal T}_i^\mu = 0$.

Regarding Lorentz invariance violation, we consider that the relation (\ref{nc}) holds in a certain orientation at the cosmic preferential referential (the one which the CMB looks maximal isotropic) and that  observer Lorentz transformations are not violated \cite{part_obs}; so the  anisotropic observables transform normally under rotations. In particular, the pressure in an arbitrary direction, given by the angles $\theta$ and $\varphi$ (usual definitions), is found from a rotation of the vacuum energy-momentum tensor  and reads
\begin{equation}
	\label{ptp}
	p(\theta, \varphi) = p_x ~ \cos^2 \theta ~ \sin^2 \varphi + p_y ~ \sin^2 \theta ~ \sin^2 \varphi + p_z ~ \cos^2 \varphi.
\end{equation}
So the mean  equation of state parameter is
\begin{equation}
	\overline \omega = \frac{  \int_0^{2\pi} \int_0^\pi p(\theta, \varphi) ~d \Omega } { {4 \pi} ~\rho} = \frac{  p_x + p_y + p_z } { 3 ~\rho}= -1.
\end{equation}

In the following section  the symmetries of the above model will be augmented by adding an axial symmetry, thus proceeding we find exact solutions and estimate the value of $B$.

\section{Axial symmetry, exact solutions, $\Lambda$CDM correspondence and the quadrupole anomaly}

Recently the Bianchi I cosmology with axial symmetry received much attention, particularly due to evidences that it provides a reasonable framework to solve  CMB  large-angle anomalies \cite{elliptical}. To implement this symmetry, we use $g = \diag \pmatrix{-1 &  a^2(t) & b^2(t) & b^2(t)}$,
\begin{equation}
	\{ \Pi^1, \Pi^2 \}_B =  ~ B ~ b
	\label{nc2}
\end{equation}
and $T_2^2 = T_3^3$, therefore
 \begin{eqnarray}
	\label{eqma1}
	2 ~\frac {\ddot b} b    + H_b^2  &=& \Lambda  - 2 ~ B_0 ~\sqrt \Lambda ~ H_b  - \frac 1k T_1^1,\\[.1in]
	\label{eqma2}
	 \frac {\ddot a} a +  \frac {\ddot b} b + H_a ~ H_b &=& \Lambda  +  B_0 ~ \sqrt \Lambda ~  H_a   - \frac 1k T_2^2, \\[.1in]
	\label{eqma3}
	 2 ~ H_a ~ H_b  + H_b^2 &=& \Lambda  - \frac 1k T_0^0.
\end{eqnarray}

For the vacuum (with $\Lambda, B \not=0$), the above equations admit a simple exact solution which reads
\begin{eqnarray}
	\label{avac}
	a &= & a_0 ~ e^{  \sqrt{ \frac {\tilde \Lambda} 3} ~ \( 1 ~ + ~  2  \tilde B_0 \) ~ t }, \\[.1in]
	\label{bvac}
	b & = &  b_0 ~ e^{  \sqrt{ \frac {\tilde \Lambda} 3} ~ \( 1 ~ - ~ \tilde B_0 \) ~ t },
\end{eqnarray}
where $\tilde \Lambda =  \Lambda ( 1 +  \frac {B_0^2} 3)$, $\tilde B_0 = \frac { B_0}{\sqrt{ 3 + B_0^2}}$ and $a_0$ and $b_0$ are constants with the suitable dimension. In the limit $B_0 \rightarrow 0$, we recover the usual de Sitter space solution ($a = b = e^{\sqrt \frac \Lambda 3 \, t}$).

The solutions (\ref{avac}, \ref{bvac}) are expected to be good approximations to the universe dynamics only for $z \apprle 0.5$. We now consider the presence of barionic and cold dark matter with matter density $\rho_M = \rho_{M0} / (a b^2)$, where $\rho_{M0}$ is a constant. The $\rho_M$ dependence on the scale factors comes from $\nabla_\mu T^\mu_0 = 0$. The exact solutions read
\begin{eqnarray}
	\label{amat}
	a(t) &=& a_0 e^{ \frac 23 B_0 ~ \sqrt \Lambda ~ (t - t_*)} ~ \sinh^{\frac 23}  \( \frac 32 ~ \sqrt{\frac {\tilde \Lambda} 3} ~ t \), \\[.1in]
	\label{bmat}
	b(t) &=& a_0 e^{ - \frac 13 B_0 ~ \sqrt \Lambda ~ (t - t_*)} ~ \sinh^{\frac 23} \( \frac 32 ~\sqrt{\frac {\tilde \Lambda} 3} ~ t \).
\end{eqnarray}
$t_*$ is an arbitrary constant. The corresponding FLRW solution is recovered  in the limit $B_0 \rightarrow 0$ [ $a = b \propto \sinh^{\frac 23} \( \frac 32 \sqrt{\frac \Lambda 3} ~ t \)$\cite{revde}].  A natural choice is to set $t_* =t_0$, the present time, so $a(t_0) = b(t_0)$. \\[.1in]

The scale factor solution for the FLRW cosmology, in the presence of a cosmological constant $\tilde \Lambda$,  describes  an average behavior of the proposed anisotropic universe in the following sense
\begin{equation}
	\label{amed}
	a_{\flrw} ( t ) = \( a ~ b^2 \)^{\frac 13} (t),
\end{equation}
hence
\begin{equation}
	H_\flrw \(  t \) = \frac 13 \( H_a(t) + 2 H_b(t) \).
\end{equation}
Thus we set the $\tilde \Lambda$ value to the experimental cosmological constant  value ($\tilde \Lambda \simeq 1.2 \times 10^{-35} ~s^{-2}$ \cite{wmap3}).

WMAP experimental results assert that $\omega = -0.967^{+0.073}_{-0.072}$ \cite{wmap3}. This imposes an upper bound for $|B_0|$ in the axial case. From the Eqs. (\ref{ptp} \ref{eqma1} \ref{eqma2} \ref{eqma3} \ref{avac} \ref{bvac}),
\begin{equation}
	\overline \omega = \frac{  \int_0^{2\pi} \int_0^\pi p(\theta, \varphi) ~d \Omega } { {4 \pi} ~\rho} = - \(1 + \frac {2 } 3 B_0^2 \),
\end{equation}
therefore  we find the  modest upper bound $ |B_0| \apprle 0.2 $ .

The universe spacial anisotropy can be quantified by the shear anisotropic parameter \cite{ shearbarrow},
\begin{equation}
	A(t) = \frac {\sigma} { \bar H} = 3 \frac {H_a - H_b} {H_a + 2 H_b} = {  3 ~ \tilde B_0 } ~ \mbox{Tanh} \( \frac 32 ~\sqrt{ \frac {\tilde \Lambda} 3} ~t \).
\end{equation}
$|A(t)|$, contrary to the usual approach to anisotropic cosmologies \cite{misner, ch, shearbarrow, cmag,  imprints}, increases with time (avoiding isotropization) and it has a maximum given by $\lim_{t \rightarrow \infty} |A(t)|=  3  ~ |{\tilde B_0}| $. Cosmologies with a similar behavior were recently studied in Ref. \cite{ans_de_eos}.

The CMB temperature anisotropies put strong bounds on anisotropic cosmologies since, at most, ${\Delta T (\theta, \phi)} /{ \bar T} \simeq 10^{-5}$.  $\bar T = 2.7 K$ is the mean measured CMB temperature, which, in the FLRW cosmology, is related with the mean original temperature  $\bar T_e$  by $\bar T = a_\flrw (t_\dec) \; \bar T_e$ (for $a_\flrw (t_0)=1$). In the model here proposed the space-time background is a source of temperature anisotropies, so  ${\Delta_B T (\theta, \phi)} / { \bar T} \apprle 10^{-5}$, where $\Delta_B$ is the variation solely due to the background. For the preferential ($x$-axis) direction,  ${\Delta_B T (0, \pi/2)} = \bar T -  a(t_\dec) \; T_x $, where $T_x$ is the original temperature of the CMB photons in that direction. We consider that the anisotropic effects were not significant for the universe dynamics by the decoupling time (in accordance with $|A(t_\dec)| \ll |A(t_0)|$), so the LSS original temperature  is almost the same in all directions.  From Eq. (\ref{amed}), the relation $\bar T = a_\flrw (t_\dec) \; \bar T_e$ also holds in the anisotropic  case as a mean value, thus we set $T_x = \bar T_e$ and find
\begin{equation}
	\frac	{\Delta_B T (0, \pi/2)}  { \bar T} = 1 -   \frac{a(t_\dec)  ~ \bar T_e}{\bar T} = 1-   \frac{a(t_\dec)  }{a_\flrw (t_\dec)}.
\end{equation}
Since the above expression should be of the same order or lesser than $10^{-5}$, we find   $|B_0| \apprle 10^{-5}$. The x-axis is the direction  with greater redshift variation from the mean one for $B_0>0$; for negative $B_0$ this upper bound analysis is done in a perpendicular direction and leads to the above bound.

\begin{figure}
\includegraphics[scale=0.8]{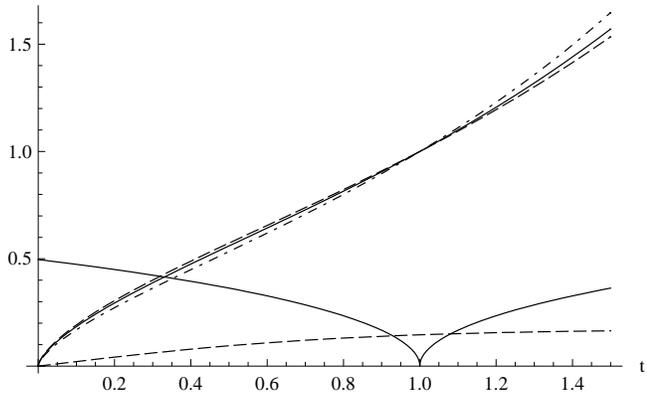}
%\vspace{-2.6 in}
\caption{The three lines that meet at $t=1$ refer to the following scale factors: $a$ (dot-dashed), $b$ (dashed) and $a_\flrw$ (solid). The other solid line is the eccentricity ($\mathsf{e} = \sqrt{1 - (a/b)^2}$ for $t \le 1$ and $\mathsf{e} = \sqrt{1 - (b/a)^2}$ for $ t \ge 1$). The other dashed line is $A = \sigma / \bar H$. Plotted with $\tilde \Lambda = 1.2 \times 10^{-35} s^{-2}$, $t_0 = 4.3 \times 10^{17} s$, $B_0 = 0.1$, $a(t_0) = b(t_0) = 1$. The horizontal axis is the time in $t_0$ unities. }
%\label{fig:}
\end{figure}

Recently it was shown that a non-null eccentricity solves the quadrupole power anomaly of the CMB spectrum  if the eccentricity of the last scattering surface (LSS) is $\mbox{e}_{LSS}  \simeq  6.4 \times 10^{-3}$ \cite{elliptical} (considering a system of coordinates with null eccentricity at present time). The present model describes a universe with eccentricity given by
\begin{eqnarray}
	&&\mathsf{e}(t, t_*) = \\
	&& \hspace{-.1in} \left \{ \hspace{-.05in} \begin{array}{ll} \sqrt{1 - \( \frac a b \)^2 } = \sqrt{1 - e^{2 B_0 \sqrt \Lambda ~ (t - t_*)}}, & \mbox{\small for $B_0 (t - t_*) \le 0$}  \\[0.2in]
	 \sqrt{1 -  \( \frac b a \)^2 } = \sqrt{1 - e^{2 B_0 \sqrt \Lambda ~ (t_* - t)}}, &  \mbox{\small for $B_0 (t - t_*) \ge 0$} \end{array} \right. \nonumber
\end{eqnarray}
Within the approach of \cite{elliptical} we set $t_* =t_0$, where $t_0$ is the age of the universe, and find  $|B_0| \simeq 1.4 \times 10^{-5}$. This value is in agreement with the previous upper bounds values. In accordance with Ref. \cite{elliptical}, ${\Delta_B T (\theta, \phi)} /{ \bar T} \lesssim 10^{-5}$ leads to the upper bound on the eccentricity, $\mathsf{e} \lesssim 10^{-2}$. Using this eccentricity value, the same bound $|B_0| \apprle 10^{-5}$ is found.

\vspace{.1in}

To second order on the eccentricity, the background alone induces the following  contribution\footnote{The  $1/3$ that appears here eliminates the monopole contribution and fixes in what directions $\frac {\Delta_B T }{\bar T}$ is null. This term is deduced in the Appendix and is in conformity with the previous $\frac {\Delta_B T }{\bar T}$ analysis, $ 1-   \frac{a  }{a_\flrw } = 1 - \( \frac a b \)^{\frac 23} = 1 - (1 - \mathsf{e}^2)^{\frac 13} \approx \frac 13 \mathsf{e}^2$. }  \cite{elliptical, moving},
\begin{equation}
	\label{gtr}
			\frac {\Delta_B T }{\bar T} (\mathbf n) =  \frac 12 \mathsf{e}^2_{\mbox{\tiny dec}} \(  n_A^2(\mathbf n) - \frac 13  \),
\end{equation}
where $\mathsf{e}^2_{\mbox{\tiny dec}} = 1 - \( \frac {a(t_{\mbox{\tiny dec}})}{ b(t{\mbox{\tiny dec}})} \)^2$ and  $n_A$ is the projection of $\mathbf n $ into the axis of symmetry. In the simplest case in which the axis of symmetry coincides with the $x$-axis, we write $n_A = \cos \Theta$, so $\Delta_B T(\mathbf n)$ depends only on $\Theta$. It is easy to see that $\frac {\Delta_B T }{\bar T} (\mathbf n) \propto Y_2^0(\mathbf n)$, implying that this elliptical background, to first order on $\mathsf{e}^2$,  modifies the quadrupole but no other multipole moment. For a coordinate system not aligned with the axis of symmetry, $\frac {\Delta_B T }{\bar T} (\mathbf n)$ is  written as a linear combination of $Y_2^m$, thus the previous conclusion remains valid.

In order  to evaluate the influence of the elliptical background to the total temperature fluctuations $\Delta T $, following  \cite{elliptical, moving}, we employ\cite{ferreira}
\begin{equation}
	\label{ans_sta}
	\Delta T = \Delta_B T + \Delta_I T,
\end{equation} 
where  $\Delta_I$ refers to the usual statistically isotropic contribution (i.e., it satisfies $\langle a^I_{lm} a^{I *}_{l'm'} \rangle = C_l \delta_{ll'} \delta_{m m'}$). Consequently, $a_{lm} = a^B_{lm} + a^I_{lm}$ and the total multipole power $Q^2_l \propto \sum_m | a_{lm}|^2$ differs from the statistically isotropic one only for $l = 2$. 

This  exclusive quadrupole modification, as implied by (\ref{gtr}, \ref{ans_sta}), guarantees no changes to the power of others multipoles. This is  welcome since  others large-angle multipole powers found from the $\Lambda$CDM model have good agreement with experiment. On the other hand, it does not help on the understanding of another CMB quadrupole anomaly, namely its alignment with the octopole\cite{zaldarriaga}.  Ref. \cite{zaldarriaga} introduces a vector $\mathbf{n}_l$ for each multipole as the direction $\mathbf{r}$ in which  $ \sum_m m^2 |a_{lm}(\mathbf{r})|^2$ is maximum\footnote{$a_{lm}(\mathbf{r})$ refers to the observed $a_{lm}$ multipole moment computed with the $z$ axis in the $\mathbf{r}$ direction.}. It was shown that $\mathbf{n}_2 \cdot \mathbf{n}_3 = 0.98$ (see also \cite{multve, axisofevil}).

The above presented results for an elliptical universe induced by an anisotropic cosmological constant rely on the first order on $\mathsf{e}^2$ approximation, so \textit{a priori} there is hope of solving the alignment if higher powers on the eccentricity are considered. The problem with this approach is that $\mathsf{e}^2_{\mbox{\tiny LSS}} \sim 10^{-5}$ and the alignment is evident from the first digits of $a_{3m}$ and $a_{2m}$, so a new very small contribution is not helpful. In the Appendix we comment more on the extension of the previous analyses to higher eccentricity powers.

\section{Conclusion}

Currently it is unclear whether  the CMB large-angle anomalies originate from some unknown systematic error (present in both the COBE and WMAP data) or if they have a physical origin. The latter points toward new physics and, in particular, to small deviations of the $\Lambda$CDM model such that statistical isotropy is violated \cite{48pol, multve, NS, axisofevil, elliptical, axial_evidence}. In this work we proposed a $\Lambda$CDM model extension whose dark energy component preserves its nondynamical character $\dot \rho_{\mbox{\tiny DE}} = 0$ but wield a vacuum anisotropic pressure. We presented exact solutions for the cosmological scale factors (\ref{avac}, \ref{bvac}, \ref{amat},  \ref{bmat}); upper bounds on the anisotropy magnitude $B_0$, considering the experimental vacuum EoS parameter and the smallness of the CMB temperature anisotropies ($\frac {\Delta T (\varphi, \theta)}T \lesssim 10^{-5} $); and lastly the $B_0$ value was estimated to be $|B_0| \simeq 1.4 \times 10^{-5}$, consistently with the previous bounds and in accordance with the quadrupole anomaly solution in an elliptical universe \cite{elliptical}.

The axial form of the  anisotropic cosmological constant here proposed, to second order on the eccentricity, only modifies the quadrupole coefficients and is capable of improving the experimental concordance for the quadrupole power. For higher orders on the eccentricity, as commented in the Appendix, the influence to others multipoles is model dependent, but such extension cannot induce the observed quadrupole-octopole alignment\footnote{Nonperturbative effects are not considered.}. In particular, for the presented model, the dipole and the octopole deviations from the $\Lambda$CDM predictions are exactly null. It is difficult to conceive a single physical mechanism capable of solving all the CMB anomalies, imprints of physics beyond the $\Lambda$CDM model probably are related with a fraction of them. 

The presented cosmological model can be achieved by two equivalent ways from a Bianchi I cosmology with cosmological constant $\Lambda$: $i$) by considering the most straightforward implementation of anisotropic vacuum pressure consistent with energy-momentum tensor conservation (\ref{div0}, \ref{amomenta}); $ii$) by implementing a poisson structure deformation between canonical momenta such that rescaling of the scale factors is not violated (\ref{nc}, \ref{nc2}). The latter procedure can be pictured as a kind of IR analogue of the scale factor noncommutativity \cite{scalefactornc}, it has already been explored in other contexts  with constant deformation parameter \cite{infrarednc}.

Bianchi cosmologies in the presence of a positive cosmological constant usually isotropize themselves with the universe expansion. This property was originally welcomed and motivated an interesting discussion on the naturalness of the universe initial conditions \cite{misner, ch}, in particular it opened the possibility that the universe has started under more general conditions   than those given by the FLRW cosmology and latter evolved to it. However, as above commented, imprints of some anisotropic (Bianchi) cosmology were found in the CMB spectrum; thus, \emph{a priori}, being these of true cosmological nature, inflation and/or dark energy need to be fine tuned to do not wipe out the anisotropic imprints. This issue emerge for instance when considering primordial magnetic fields as a source of current cosmological anisotropies \cite{cmag, elliptical}. Here we  employed an anisotropic model outside the ``FLRW equivalent class" (i.e., not in conformity with the cosmic no-hair theorem conditions \cite{b_isotropization-w}) and it was shown that such model is viable and can also solve the quadrupole power anomaly. While this paper was being prepared, some similar ideas appeared in Ref. \cite{ans_de_eos}.

This recently explored approach in which the universe anisotropy after inflation  increases with time (e.g., \cite{moving, ans_de_eos}) is not only consistent with observations, it  also improves the concordance with experiment and suggest other interpretations for dark energy. New phenomenological anisotropic alternatives for the dark energy modeling might, we hope, inspire and lead to more fundamental models.

\vspace{.2in}
\noindent
{ \bf Acknowledgments}

The author would like to thank Jorge Gamboa for suggesting the application of an IR noncommutativity to cosmology, for useful discussions and for the kind reception in Chile. The author also thanks Fernando M\'endez, Cl\'ovis Wotzasek and Marcelo S. Guimar\~aes for useful discussions. This work was supported by FONDECYT-Chile grant n. 3070008.

\appendix

\section{On eccentricity perturbation}

Here we derive Eq.(\ref{gtr}) and comment on the background influence to others multipoles if  higher eccentricity powers are considered. The axisymmetric Bianchi I line element can be  written in terms of an anisotropic perturbation $h_{ij}$ and a conformal scale factor $\alpha (\eta)$ as
\begin{equation}
	ds^2 =  \alpha^2(\eta) \[ - d\eta^2 + \(\delta_{ij} + h_{ij} \) dx^i dx^j \].
\end{equation}

Considering (\ref{amed}), we set $\alpha  = (a b^2)^{\frac 13}$,  thus
\begin{eqnarray}
	&& h_{11} = \( \frac a \alpha \)^2 - 1 = \( \frac a b \)^{\frac 43} -1 \\ 
	&&  ~~~~~~~ = - \frac 23 \mathsf{e}^2 - \frac 1 9 \mathsf{e}^4 + O(\mathsf{e}^6), \nonumber \\ [.1in]
	&& h_{22} = h_{33} =  \( \frac b a \)^{\frac 23} -1 	\\ 
	&& ~~~~~~~ = \frac 13 \mathsf{e}^2 + \frac 29 \mathsf{e}^4 + O(\mathsf{e}^6), \nonumber
\end{eqnarray}
with $\mathsf{e}^2 \equiv 1 - \frac {a^2}{b^2}$.

	Let $h_{ij} = h_{ij}^{(1)} + h_{ij}^{(2)} +... ~$,  with $h_{ij}^{(k)} \propto \mathsf{e}^{2 k } $. $h_{ij}^{(1)}$ is traceless (and therefore  in accordance with the standard tensorial perturbation theory \cite{giovannini}). Some redefinitions are needed in order to eliminate the traceful part of\footnote{Up to $\mathsf e^4$, they read: $\alpha^2 = (a b^2)^{\frac 23}  ~ ( 1 + \frac 19 \mathsf{ e}^4)$,  $h_{11}^{(2)} = - \frac 2 9 \mathsf e^4$, $h_{22}^{(2)} = h_{33}^{(2)}= \frac 1 9 \mathsf e^4$. The $\alpha$ redefinition is necessary to preserve $ds^2$. $(a b^2)^{\frac 13}$ indeed works as a good mean scale factor, but, as implied by Eq. (\ref{amed}), it is not an exact relation, a small time (or $\Lambda$)  redefinition is also necessary} $h_{ij}^{(2)}$.

	The background temperature fluctuations come from the tensor Sachs-Wolfe effect, which to first order reads \cite{giovannini}
\begin{eqnarray}
	\frac {\Delta_B T }{\bar T} (\mathbf n) && =  \frac 12 \int_{\eta_\dec}^{\eta_0} \frac{d h^{(1)}_{ij}}{d \eta} n^i n^j ~ d\eta  \\
	&& = \frac {\mathsf{e}^2_\dec} 2 \(  \frac 23 (n^1)^2 - \frac 13 (n^2)^2 - \frac 13 (n^3)^2 \),  \nonumber
\end{eqnarray}
with $\mathbf n \cdot \mathbf n = 1$, and leads to Eq.(\ref{gtr}) apart from the time gauge fixing.

In the following we analyze the background influence to the dipole and octopole if higher eccentricity powers are considered. In fact, the background alone contribution  to these multipoles is exactly null.  To any order on the eccentricity, consider  a system of coordinates in which the $x$-axis coincides with the preferential direction, hence  $\frac {\Delta_B T}{\bar T}$ has even parity symmetry  in the interval   $\Theta \in [0,\pi]$ (induced by the exact elliptical symmetry) and one concludes that\footnote{This can also be found directly from the redshift formulas of Refs. \cite{moving, ans_de_eos}.}
\begin{equation}
	\label{gty}
	a^B_{lm} = \int Y^{m  *}_l ~ \frac {\Delta_B T}{\bar T} ~ d\Omega = 0 ~~~~~~~ \forall \mbox{ odd } l.
\end{equation}  
Rotations do not change this conclusion.

Any odd $l$ modification in the multipole moments induced by the background comes from the Eq. (\ref{ans_sta}) extension to higher eccentricity powers, which depends on how the Gaussian fluctuations interact with the anisotropic perturbation. In the presented approach, since the source of  anisotropies is non-dynamical, the linearity of Eq. (\ref{ans_sta}) is preserved and, in particular, there is no change to the dipole or the octopole. 

One may extend the presented approach in order to include the possibility of dark energy perturbations. In that case, since both $\frac {\Delta_I T}{\bar T}$ and $\frac {\Delta_B T}{\bar T}$ are roughly of the same order, the interaction should be properly found by an evaluation of a second order Sachs-Wolfe effect that considers both the anisotropic and the isotropic perturbations together. The second order Sachs-Wolfe effect of the \asp isotropic" part alone is under current research, in particular it induces non-Gaussianity (e.g., \cite{2sw}).  The details of this issue are outside this paper purposes, we only add that, up to $\mathsf e^4$ (second order), the Eq. (\ref{gtr}) should have the \textit{aspect} $\Delta T^{(1,2)} = \Delta_B T^{(1,2)} + \Delta_I T^{(1,2)} + \Delta_B T^{(1)} \Delta_I T^{(1)}$, leading to $a^{(1,2)}_{3 m} = a^{I  ~ (1,2)}_{3 m} + \mathsf e^2 \sum_{l' m'} a^{I(1)}_{l'm'}\int Y_3^{m*} Y_2^0 Y_{l'}^{m'} d\Omega $. Hence,  background contributions to the dipole and octopole are expected if the mentioned interaction is added\footnote{From the Clebsch-Gordon coefficients, one sees that the background contribution to $a^{(1,2)}_{l m}$ becomes null for $l>5$, generating no contribution to the CMB microphysics, as expected.}. This tiny influence is not related with the quadrupole-octopole alignment, nonetheless  future high precision CMB observations should be sensible to such influence.

\end{document}